\documentclass[10pt]{article}
\usepackage{colm2024_conference}
\usepackage{microtype}
\usepackage{xspace}
\usepackage{booktabs}
\usepackage{multirow}
\usepackage{amsmath,amsfonts,amssymb}
\usepackage{graphicx}
\usepackage{url}
\usepackage{textcomp}
\usepackage{xcolor}
\usepackage[labelfont=bf]{caption}
\usepackage[T1]{fontenc}
\usepackage{array}
\usepackage{listings}
\usepackage{mdframed}
\usepackage{tikz}
\usetikzlibrary{arrows.meta}
\usepackage{pgfplots}
\usepgfplotslibrary{groupplots}
 \usepackage{listings}
 \usepackage{xcolor}

\usepackage{tikz}
\usetikzlibrary{arrows.meta,positioning}

 \definecolor{precbg}{RGB}{214,230,248}   %
 \definecolor{intentbg}{RGB}{253,224,196} %
 \definecolor{codebg}{RGB}{248,248,248}
 \definecolor{codekw}{RGB}{0,62,138}
 \definecolor{codestr}{RGB}{25,110,45}

 \newcommand{\hlprec}[1]{{\setlength{\fboxsep}{1pt}\colorbox{precbg}{\ttfamily #1}}}
 \newcommand{\hlintent}[1]{{\setlength{\fboxsep}{1pt}\colorbox{intentbg}{\ttfamily #1}}}

 \usepackage[most]{tcolorbox}

 \lstdefinestyle{motiv}{
   language=Python,
   basicstyle=\ttfamily\scriptsize,
   keywordstyle=\color{codekw}\bfseries,
   stringstyle=\color{codestr},
   numbers=left, numberstyle=\tiny\color{gray}, numbersep=4pt,
   xleftmargin=1.4em, framexleftmargin=1.4em,
   frame=tb, framerule=0.4pt, rulecolor=\color{gray!50},
   backgroundcolor=\color{codebg},
   columns=fullflexible, keepspaces=true, showstringspaces=false,
   escapeinside={(*@}{@*)},
   aboveskip=2pt, belowskip=2pt
 }

\definecolor{stage1}{HTML}{0072B2}        %
\definecolor{stage2}{HTML}{D55E00}        %
\definecolor{stage3}{HTML}{0E8A63}        %
\definecolor{darktext}{HTML}{1F2328}      %
\definecolor{greytext}{HTML}{5B6470}      %
\definecolor{fgbaseline}{RGB}{214,230,248}    %
\definecolor{fgalternative}{RGB}{240,150,70}  %
\definecolor{fgaltlight}{RGB}{253,224,196}   %
\definecolor{fgSonnet}{HTML}{0072B2}
\definecolor{fgOpus}{HTML}{56B4E9}
\definecolor{fgGfive}{HTML}{D55E00}
\definecolor{fgGsix}{HTML}{E69F00}

\newcommand{\codefontsmall}[1]{{\small\texttt{#1}}}

\newcommand{\rqanswer}[2]{}
\newsavebox{\motivbox}
\usepackage{fontawesome5}
\newsavebox{\figbox}

\definecolor{SemNavy}{HTML}{213C55}
\definecolor{SemTeal}{HTML}{187C80}
\definecolor{SemMuted}{HTML}{637786}
\definecolor{SemPale}{HTML}{F0F6F7}
\definecolor{SemLine}{HTML}{CDDDE2}
\hypersetup{colorlinks=true,linkcolor=SemNavy,citecolor=SemTeal,urlcolor=SemTeal,
 pdfauthor={Haichao Zhu, Geyang Xu, Zhaorui Yang, Jiyuan Wang, Qian Zhang},
 pdftitle={Precedent or Intent? What Coding Agents Follow When Evolving Existing Code}}
\setcitestyle{numbers,square,comma,sort&compress}
\tcbuselibrary{most}
\newtcolorbox{abstractbox}{enhanced,breakable,colback=SemPale,colframe=SemLine,
 boxrule=0.4pt,arc=1.5mm,left=10pt,right=10pt,top=5pt,bottom=5pt,
 borderline west={2pt}{0pt}{SemTeal},before skip=8pt,after skip=12pt}
\renewenvironment{abstract}{\noindent{\color{SemNavy}\bfseries Abstract}\par\small}{\par}
\renewcommand{\rqanswer}[2]{\begin{mdframed}[backgroundcolor=SemPale,linecolor=SemTeal,
 linewidth=0.8pt,topline=false,rightline=false,bottomline=false,
 innerleftmargin=9pt,innerrightmargin=9pt,innertopmargin=6pt,innerbottommargin=6pt,
 skipabove=8pt,skipbelow=6pt,nobreak=true]\textbf{\color{SemTeal}#1.} #2\end{mdframed}}
\makeatletter
\renewcommand{\headrulewidth}{0.5pt}
\renewcommand{\headrule}{{\color{SemLine}\hrule\@height\headrulewidth\@width\headwidth\vskip-\headrulewidth}}
\fancypagestyle{firstpage}{\fancyhf{}
 \fancyhead[L]{\includegraphics[height=23pt]{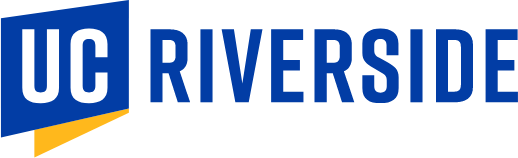}\hspace{12pt}\includegraphics[height=23pt]{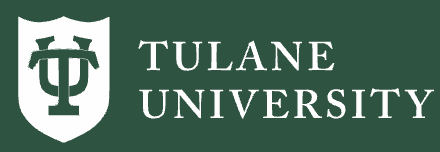}}
 \fancyhead[R]{\small\color{SemMuted} October 26, 2026}
 \fancyfoot[C]{\small\color{SemMuted}\thepage}}
\renewcommand{\@maketitle}{\vspace*{7pt}\begingroup\centering
 {\fontsize{17}{21}\selectfont\bfseries\color{SemNavy}\@title\par}
 \vspace{12pt}{\normalsize\@author\par}\endgroup\vspace{5pt}\thispagestyle{firstpage}}
\makeatother
\pgfplotsset{compat=1.17}
\title{Precedent or Intent?\\ What Coding Agents Follow When Evolving Existing Code}
\author{%
\textbf{Haichao Zhu}$^{2}$\quad
\textbf{Geyang Xu}$^{1,*}$\quad
\textbf{Zhaorui Yang}$^{1}$\quad
\textbf{Jiyuan Wang}$^{3}$\quad
\textbf{Qian Zhang}$^{1}$\\[7pt]
{\small $^{1}$University of California, Riverside\quad $^{2}$Independent Researcher\quad $^{3}$Tulane University}\\[5pt]
}
\date{}

\begin{document}

\maketitle

\begingroup
\renewcommand{\thefootnote}{*}
\footnotetext{Corresponding author:
  \href{mailto:gxu048@ucr.edu}{gxu048@ucr.edu}.}
\endgroup

\begin{abstractbox}\begin{abstract}Existing code reveals past implementation decisions but rarely explains the reasoning behind them. When coding agents implement new features or modify existing functionality, past decisions may conflict with new task instructions. This raises a fundamental question for agentic software evolution: when multiple implementations satisfy the same functional requirements, how do coding agents decide whether to follow existing code or new task instructions?

\medskip

We investigate this question through bidirectional, controlled experiments in purpose-built repositories, followed by validation in real-world open-source repositories. Our approach isolates three competing sources of guidance: (1) Agents' Default Preferences, measured without repository precedents; (2) Repository Precedents, examined by manipulating the frequency and location of existing code examples; and (3) Explicit Task Instructions, tested by introducing rules that conflict with existing code.

\medskip

We conduct 5,625 trials across four coding agents and six implementation pairs, with validation in four open-source repositories spanning three programming languages. We find that agents share the same default implementation choices, yet overwhelmingly follow repository precedent when it conflicts with those defaults (359/360 trials). Crucially, not all precedents carry equal influence: three supporting local files yield 96\% adoption, compared with only 4\% for sibling-resource files, and changing file placement alone can shift adoption from 0\% to 100\% at a fixed example count. When new task intent conflicts with existing code, written priority rules achieve substantially higher success than code examples alone (59/60 versus 38/60 joint successes for two agents on two implementation pairs). The effectiveness of explicit rules also extends to open-source repositories, where a harness-loaded rule raises joint success from 0/120 to 117/120. These findings motivate design-decision traceability that preserves not
only implementation choices but also the rationale and priorities that
should guide future changes.

\end{abstract}\end{abstractbox}

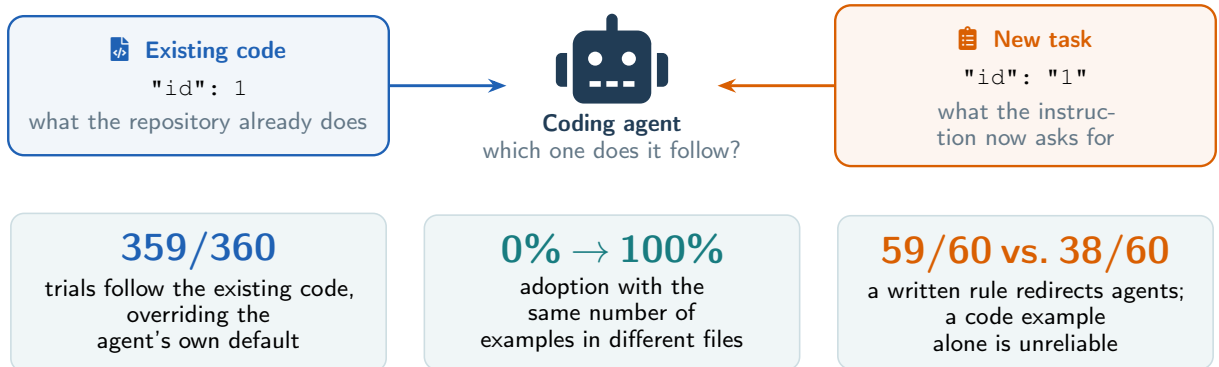
\begin{figure*}[!h]
\centering
\definecolor{tzBlue}{RGB}{31,97,180}
\definecolor{tzOrange}{RGB}{217,95,2}
\resizebox{\textwidth}{!}{%
\begin{tikzpicture}[
  font=\sffamily\small,
  card/.style={draw=#1, line width=0.9pt, rounded corners=6pt, fill=#1!6, inner sep=8pt, text width=4.5cm, align=center},
  stat/.style={draw=SemLine, line width=0.6pt, rounded corners=5pt, fill=SemPale, inner sep=7pt, text width=4.5cm, align=center},
  arr/.style={-{Stealth[length=7pt]}, line width=1.1pt}]
\node[card=tzBlue] (prec) at (-5.5,0) {\textbf{\color{tzBlue}\faFileCode~~Existing code}\\[4pt]{\ttfamily\footnotesize "id": 1}\\[3pt]{\footnotesize\color{SemMuted}what the repository already does}};
\node[card=tzOrange] (intent) at (5.5,0) {\textbf{\color{tzOrange}\faClipboardList~~New task}\\[4pt]{\ttfamily\footnotesize "id": "1"}\\[3pt]{\footnotesize\color{SemMuted}what the instruction now asks for}};
\node[align=center] (agent) at (0,0) {{\fontsize{34}{34}\selectfont\color{SemNavy}\faRobot}\\[3pt]\textbf{\color{SemNavy}Coding agent}\\{\footnotesize\color{SemMuted}which one does it follow?}};
\draw[arr, tzBlue] (prec.east) -- ++(1.55,0);
\draw[arr, tzOrange] (intent.west) -- ++(-1.55,0);
\node[stat] (s1) at (-5.5,-2.75) {{\Large\bfseries\color{tzBlue}359/360}\\[2pt]{\footnotesize trials follow the existing code,\\ overriding the agent's own default}};
\node[stat] (s2) at (0,-2.75) {{\Large\bfseries\color{SemTeal}0\% $\rightarrow$ 100\%}\\[2pt]{\footnotesize adoption with the same number of\\ examples in different files}};
\node[stat] (s3) at (5.5,-2.75) {{\Large\bfseries\color{tzOrange}59/60 vs.\ 38/60}\\[2pt]{\footnotesize a written rule redirects agents;\\ a code example alone is unreliable}};
\end{tikzpicture}}
\caption{Coding agents that extend a repository are pulled between the implementation already in the code and the intent of the new task. In controlled experiments with four agents and six implementation pairs, existing code wins by default, and a written rule is what reliably changes course.}
\label{fig:teaser}
\end{figure*}

\section{Introduction}\label{sec:introduction}
Software development involves the continuous evolution of existing systems. As LLM-based coding agents take on increasingly complex software engineering tasks, their role is expanding beyond individual issue resolution to implementing changes across evolving repositories~\cite{yang2024sweagent,jimenez2024swebench,zhu2026nitr,deng2026evoclaw,le2026sweevobenchmarkingcodingagents}. Unlike building from scratch, extending an existing repository means working inside decisions made during earlier development, whose rationale is often lost. These decisions influence how new functionality is implemented, even when multiple alternatives satisfy the same functional requirements~\cite{deng2026evoclaw,zhu2026nitr}. However, existing code rarely explains why a particular implementation was chosen or whether that decision should be preserved as requirements evolve. Coding agents must therefore reconcile past implementation decisions with the intent expressed in new task specifications.

\vspace{.3em}
\noindent{\bf Past Decisions vs. New Intent.}
Existing code records past implementation decisions. {\em Every new change inherits decisions made in the past, but how does a coding agent determine which decisions to preserve, particularly when a new task expresses a different intent?} This question raises two challenges in agentic software evolution.

First, the rationale behind past implementation decisions is often unavailable. Existing code reveals what previous developers chose, but rarely why~\cite{liang2023implementation,zhao2024drminer,sharma2021rationale}. For example, a repository may use an if-else chain rather than a dictionary lookup because of an earlier requirement or simply a developer's preference. Without this context, a coding agent cannot determine whether the existing code reflects a constraint that should be preserved.

Second, past implementation decisions may conflict with the new task intent. Different parts of a repository may follow different implementation patterns, while new task instructions may favor an alternative~\cite{wang2025style,li2025crossfile}. For example, an agent adding an API endpoint may encounter existing handlers that serialize identifiers as integers, while the new task may specify a preference for string identifiers. Both options may satisfy functional requirements, but represent different implementation choices. The agent must determine whether to follow the existing precedent or the new instructions. 

Despite substantial progress in evaluating coding agents in long-horizon software engineering tasks~\cite{jimenez2024swebench,le2026sweevobenchmarkingcodingagents,zhu2026nitr}, how agents prioritize past implementation decisions and new task instructions remains poorly understood. In particular, we lack systematic evidence on the relative influence of {\em agents' default preferences, implementation patterns in task-related code and elsewhere in the repository, and explicit task guidance.}

\vspace{.3em}
\noindent{\bf Approach.}
We investigate how coding agents prioritize competing sources of guidance when modifying existing software repositories. We examine three factors: (1) agents' default implementation preferences, (2) repository precedent established by past implementations, and (3) explicit guidance in new task specifications. Our approach combines bidirectional controlled experiments in purpose-built repositories with a validation study in real-world open-source repositories.

We first establish agents' default preferences in a manually constructed Python/Flask repository. We design six implementation-choice cases that span naming, data representation, error handling, and control flow, each allowing two functionally correct alternatives, such as integer versus string identifiers. By removing relevant repository examples, we establish a baseline for two complementary controlled interventions:

\begin{itemize}
    \item {\em Repository Precedent.} We examine how historical implementations shift agents' default choices.
    We distinguish local precedent, established by existing endpoints for the resource being extended, from sibling-resource precedent, established by endpoints for other resources in the same repository. Holding task requirements fixed, we vary the number and placement of supporting examples and introduce conflicting patterns across resources. These interventions isolate how the prevalence and location of existing patterns influence agents’ choices.

    \item {\em Explicit Task Guidance.} We examine whether new instructions can override established precedent. Starting with code that favors one implementation, we compare four guidance conditions: no guidance, a written priority rule, a code example favoring the alternative, and both combined. We further test whether a rule reaches the agent equally well through the task description or through the harness's native instruction file. This reveals how different forms and channels of guidance redirect agents toward a new task intent.
\end{itemize}

\vspace{.3em}
\noindent{\bf Key Findings.}
Our study comprises 5,625 trials across six implementation pairs and four models: Claude Sonnet-4-6, Claude Opus-4-8, Codex gpt-5.6-sol, and Codex gpt-6-sol. We further evaluate four open-source repositories spanning Python, Go, and TypeScript: FastAPI RealWorld~\cite{repo-fastapi-realworld}, FastAPI Clean Example~\cite{repo-fastapi-clean}, Go REST API~\cite{repo-go-rest}, and ExpressJS API~\cite{repo-expressjs-api}. Three key findings emerge, highlighting the need to make implementation decisions and their intended priority explicit during software evolution.

First, {\it repository precedent overturns agents' shared default choices in 359/360 trials.}
Without relevant repository precedents, all four agents favor the same implementation in each of the six pairs. Yet when all existing endpoints consistently use the opposite implementation, agents follow that precedent in 359/360 trials, despite unchanged functional requirements. In four open-source repositories, agents follow local precedent in all 945 functionally successful, classifiable error-handling implementations and 355 of 417 (85\%) functionally successful response-envelope implementations.

Second, \emph{a single well-placed example can outweigh multiple examples elsewhere.}
With three supporting files, local precedents yield 96\% adoption (347/360), compared with only 4\% (13/360) for sibling-resource precedents. Even within local code, changing file placement while holding the number of examples fixed can shift adoption from 0\% to 100\%. In one setting, a single well-placed example even outperforms two examples in other files. These findings show that the influence of repository precedent cannot be explained by example frequency alone.

Third, \emph{written priority rules override conflicting precedent more reliably than code examples alone (98\% vs.\ 63\%).}
When all existing endpoints favor the opposite implementation, written priority rules achieve both functional success and target adoption in 59/60 trials, compared with 38/60 for code examples alone and 0/60 without guidance. The effectiveness of written rules also extends to four open-source repositories, where a rule in the harness's native instruction file raises joint success from 0/120 to 117/120. 

Figure~\ref{fig:teaser} summarizes these findings. This paper makes three contributions:
\begin{itemize}
    \item We establish the prioritization of historical code and new task intent as a research problem in software evolution, introducing bidirectional controlled experiments to isolate their influence on coding agents.

    \item We reveal that local precedent overrides default preferences and that where examples are placed can change adoption from 0\% to 100\% at a fixed example count.

    \item We demonstrate that explicit rules override conflicting repository precedent more reliably than examples, motivating design-decision traceability to preserve implementation intent and \mbox{priorities}.
\end{itemize}

\section{Motivating Example}\label{sec:motivation}
\begin{figure*}[t]
\centering\begin{lrbox}{\motivbox}\begin{minipage}{7.6in}
\centering
\begin{minipage}[t]{0.36\linewidth}
\noindent\raggedright\textbf{\footnotesize (a) Existing local handler: \texttt{tasks\_list.py}}\par\vspace{2pt}
\begin{lstlisting}[style=motiv]
@bp.get("/projects/<int:pid>/tasks")
def list_tasks(pid):
    tasks = Task.query.filter_by(
        project_id=pid).all()
    return jsonify([
        {"id": (*@\hlprec{t.id}@*), (*@\label{line:a-id}@*)
         "name": t.name}
        for t in tasks
    ])
\end{lstlisting}
\end{minipage}\hfill
\begin{minipage}[t]{0.31\linewidth}
\noindent\raggedright\textbf{\footnotesize (b) New handler: follows precedent}\par\vspace{2pt}
\begin{lstlisting}[style=motiv]
@bp.get("/projects/<int:pid>" (*@\label{line:b-route}@*)
        "/tasks/<int:tid>") (*@\label{line:b-route2}@*)
def get_task(pid, tid):
    task = Task.query.filter_by( (*@\label{line:b-query}@*)
        id=tid, project_id=pid
    ).first_or_404() (*@\label{line:b-query-end}@*)
    return jsonify({
        "id": (*@\hlprec{task.id}@*), (*@\label{line:b-id}@*)
        "name": task.name
    })
\end{lstlisting}
\end{minipage}\hfill
\begin{minipage}[t]{0.31\linewidth}
\noindent\raggedright\textbf{\footnotesize (c) New handler: follows new intent}\par\vspace{2pt}
\begin{lstlisting}[style=motiv]
@bp.get("/projects/<int:pid>"
        "/tasks/<int:tid>")
def get_task(pid, tid):
    task = Task.query.filter_by(
        id=tid, project_id=pid
    ).first_or_404()
    return jsonify({
        "id": (*@\hlintent{str(task.id)}@*), (*@\label{line:c-id}@*)
        "name": task.name
    })
\end{lstlisting}
\end{minipage}
\end{minipage}\end{lrbox}\resizebox{\textwidth}{!}{\usebox{\motivbox}}
\caption{Illustrative Flask handlers for the motivating example. (a)~An existing local handler serializes task IDs as numbers (line~\ref{line:a-id}). (b) and (c)~Two implementations of the new endpoint retrieve the same task with an identical query (lines~\ref{line:b-query}--\ref{line:b-query-end}) and differ only in how they serialize the identifier (line~\ref{line:b-id}). (b)~carries over the repository precedent (\colorbox{precbg}{\strut blue}), whereas (c)~follows a new instruction to serialize identifiers as strings (\colorbox{intentbg}{\strut orange}).}
\label{fig:motivation-code}
\end{figure*}

Consider a coding agent asked to add the endpoint \texttt{GET /projects/\textless id\textgreater/tasks/\textless id\textgreater} to an existing Flask API in Figure~\ref{fig:motivation-code}. The endpoint retrieves a task and returns its information as JSON. Although the required behavioral specification is given, some concrete implementation decisions remain open. For example, the task identifier may be serialized as a JSON number (Figure~\ref{fig:motivation-code}(b)) or as a string (Figure~\ref{fig:motivation-code}(c)). %

\vspace{.3em}
\noindent{\bf Past Decisions as Precedent.}
The existing task handler in Figure~\ref{fig:motivation-code}(a) serializes the identifiers as numbers using \texttt{task.id}. Such existing implementations establish \emph{repository precedents}: past implementation decisions that may guide subsequent changes. An agent following this precedent would produce Figure~\ref{fig:motivation-code}(b). However, a repository does not necessarily follow a single implementation pattern across locations within a codebase. The two ID representations illustrated in Figure~\ref{fig:motivation-code}(b) and (c) may coexist in existing code, appearing in different files or even within the same file. The agent may therefore encounter conflicting historical implementations rather than a single pattern to follow.

\vspace{.3em}
\noindent{\bf New Intent.}
Following existing implementation decisions helps maintain consistency as software evolves. For example, reusing the numeric-ID representation in Figure~\ref{fig:motivation-code}(a) would keep the new endpoint consistent with existing task responses. 

However, new requirements may require changes to earlier design choices. Suppose the new task explicitly states: {``Serialize task identifiers as strings in the new endpoint. This instruction takes precedence over the ID representations used by existing handlers.''} The agent now \emph{faces a conflict: following the existing precedent preserves consistency but contradicts the new instruction, whereas following the instruction satisfies the new intent but departs from the established implementations.} 

This illustrates a fundamental challenge in software evolution: how coding agents choose between preserving past implementation decisions and following newly specified intent.

\vspace{.3em}
\noindent{\bf The Research Problem.}
The resulting code reveals which implementation the agent selected, but not what guided that choice. An agent may follow its default preference, adopt a precedent found in the repository, or follow an explicit task instruction. Our research investigates how these three sources of guidance influence implementation choices, particularly when historical precedents and new task intent conflict. To distinguish their effects, we systematically vary repository precedents and task guidance while keeping the functional requirements unchanged.

\section{Approach}\label{sec:approach}
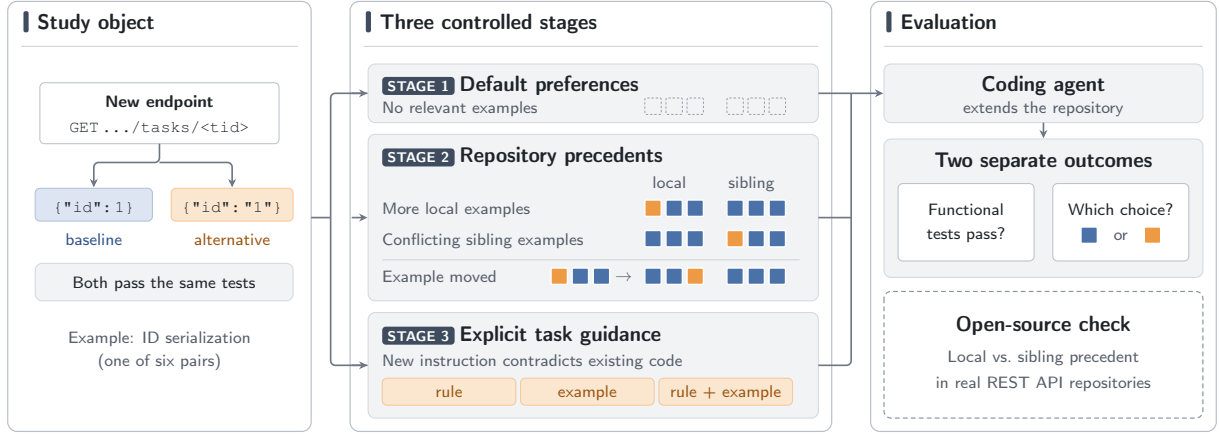
\begin{figure*}[t]
\centering
\definecolor{fgInk}{HTML}{1F2328}     %
\definecolor{fgMuted}{HTML}{555D69}   %
\definecolor{fgAccent}{HTML}{3E4C5E}  %
\definecolor{fgFrame}{HTML}{B9BFC8}   %
\definecolor{fgRule}{HTML}{D3D7DD}    %
\definecolor{fgCard}{HTML}{F1F3F5}    %
\definecolor{fgArrow}{HTML}{6E7783}   %
\definecolor{fgBase}{HTML}{4A72A8}    %
\definecolor{fgBaseFill}{HTML}{DCE5F1}%
\definecolor{fgBaseText}{HTML}{2F5486}%
\definecolor{fgAlt}{HTML}{EE9A45}     %
\definecolor{fgAltFill}{HTML}{FBE6D0}%
\definecolor{fgAltText}{HTML}{9A5512}%
\pgfmathsetlengthmacro{\fgU}{1pt}%
\def\fgH{184}%
\def\fgB{fgBase}\def\fgA{fgAlt}%
\def\fgPanel#1#2#3#4{%
  \path[draw=fgFrame,line width=0.5pt,fill=white,rounded corners=4pt]
     ({#1+0.3},0.3) rectangle ({#2-0.3},{\fgH-0.3});
  \path[fill=#3,rounded corners=0.8pt] ({#1+7},4.5) rectangle ({#1+9.2},13.5);
  \node[font=\sffamily\footnotesize\bfseries,text=fgInk,anchor=base west] at ({#1+13},12) {#4};
  \draw[fgRule,line width=0.5pt] ({#1+6},19.5) -- ({#2-6},19.5);}
\def\fgNum#1#2#3#4{\node[fill=#3,text=white,font=\sffamily\tiny\bfseries,rounded corners=1.5pt,
  inner xsep=2.3pt,inner ysep=1.7pt,anchor=base west] at (#1,#2) {STAGE #4};}
\def\fgSq#1#2#3{\path[fill=#3,draw=white,line width=0.6pt,rounded corners=0.6pt] (#1,#2) rectangle ++(7,-7);}
\def\fgSqE#1#2{\path[draw=fgArrow!70,dash pattern=on 1pt off 0.8pt,line width=0.4pt,rounded corners=0.6pt] (#1,#2) rectangle ++(7,-7);}
\def\fgTri#1#2#3#4#5{\fgSq{#1}{#2}{#3}\fgSq{#1+9}{#2}{#4}\fgSq{#1+18}{#2}{#5}}
\resizebox{\textwidth}{!}{\begin{tikzpicture}[
  x=\fgU,y=-1pt,font=\sffamily,
  every node/.style={inner sep=0pt,outer sep=0pt,anchor=base,text=fgInk},
  fgT/.style={font=\sffamily\footnotesize\bfseries,text=fgInk},
  fgS/.style={font=\sffamily\scriptsize,text=fgMuted},
  fgC/.style={font=\sffamily\scriptsize,text=fgInk},
  fgcard/.style={fill=fgCard,draw=fgRule,line width=0.4pt,rounded corners=3pt},
  fgbox/.style={fill=white,draw=fgFrame,line width=0.4pt,rounded corners=2pt},
  fgline/.style={line width=0.7pt,draw=fgArrow},
  fgmain/.style={-{Stealth[length=3.6pt,width=3.4pt]},line width=0.7pt,draw=fgArrow}]
\useasboundingbox (0,0) rectangle (516,\fgH);
\fgPanel{0}{130}{fgAccent}{Study object}
\fgPanel{146}{352}{fgAccent}{Three controlled stages}
\fgPanel{368}{516}{fgAccent}{Evaluation}

\path[fgbox] (14,35) rectangle (116,61);
\node[fgC] at (65,45.5) {\textbf{New endpoint}};
\node[fgC] at (65,56) {\texttt{GET\,.../tasks/<tid>}};
\draw[fgmain,rounded corners=2pt] (65,61) -- (65,69) -- (37,69) -- (37,79.6);
\draw[fgmain,rounded corners=2pt] (65,69) -- (96,69) -- (96,79.6);
\path[fill=fgBaseFill,draw=fgBase!45,line width=0.4pt,rounded corners=2pt] (12,80) rectangle (62,94);
\node[fgC] at (37,90) {\texttt{\{"id":\,1\}}};
\path[fill=fgAltFill,draw=fgAlt!55,line width=0.4pt,rounded corners=2pt] (70,80) rectangle (122,94);
\node[fgC] at (96,90) {\texttt{\{"id":\,"1"\}}};
\node[font=\sffamily\scriptsize,text=fgBaseText] at (37,104) {baseline};
\node[font=\sffamily\scriptsize,text=fgAltText] at (96,104) {alternative};
\path[fgcard] (12,112) rectangle (122,128);
\node[fgC] at (67,122.5) {Both pass the same tests};
\node[fgS] at (65,146) {Example: ID serialization};
\node[fgS] at (65,156) {(one of six pairs)};

\path[fgcard] (154,27) rectangle (346,52);
\fgNum{160}{38.5}{fgAccent}{1}
\node[fgT,anchor=base west] at (193,38.5) {Default preferences};
\node[fgS,anchor=base west] at (160,48) {No relevant examples};
\fgSqE{272}{48.5}\fgSqE{281}{48.5}\fgSqE{290}{48.5}
\fgSqE{307}{48.5}\fgSqE{316}{48.5}\fgSqE{325}{48.5}

\path[fgcard] (154,57) rectangle (346,128);
\fgNum{160}{69}{fgAccent}{2}
\node[fgT,anchor=base west] at (193,69) {Repository precedents};
\node[fgS] at (282.5,80) {local};
\node[fgS] at (317.5,80) {sibling};
\node[fgS,anchor=base west] at (160,91) {More local examples};
\fgTri{272}{92}{\fgA}{\fgB}{\fgB}\fgTri{307}{92}{\fgB}{\fgB}{\fgB}
\node[fgS,anchor=base west] at (160,104) {Conflicting sibling examples};
\fgTri{272}{105}{\fgB}{\fgB}{\fgB}\fgTri{307}{105}{\fgA}{\fgB}{\fgB}
\node[fgS,anchor=base west] at (160,120) {Example moved};
\fgTri{232}{121}{\fgA}{\fgB}{\fgB}
\node[fgS] at (263,119.5) {$\rightarrow$};
\fgTri{272}{121}{\fgB}{\fgB}{\fgA}\fgTri{307}{121}{\fgB}{\fgB}{\fgB}
\draw[fgRule,line width=0.5pt] (160,110.5) -- (338,110.5);

\path[fgcard] (154,133) rectangle (346,178);
\fgNum{160}{145}{fgAccent}{3}
\node[fgT,anchor=base west] at (193,145) {Explicit task guidance};
\node[fgS,anchor=base west] at (160,156) {New instruction contradicts existing code};
\foreach \x/\t in {160/rule,219/example,278/{rule + example}}{
  \path[fill=fgAltFill,draw=fgAlt!55,line width=0.4pt,rounded corners=2pt] (\x,161) rectangle ++(57,11.5);
  \node[font=\sffamily\scriptsize,text=fgAltText] at (\x+28.5,169.2) {\t};}

\path[fgcard] (374,27) rectangle (510,52);
\node[fgT] at (442,38.5) {Coding agent};
\node[fgS] at (442,48) {extends the repository};
\draw[fgmain] (442,52) -- (442,58.7);
\path[fgcard] (374,59) rectangle (510,118);
\node[fgT] at (442,71) {Two separate outcomes};
\path[fgbox] (380,78) rectangle (438,111);
\node[fgC] at (409,91.5) {Functional};
\node[fgC] at (409,101.5) {tests pass?};
\path[fgbox] (446,78) rectangle (504,111);
\node[fgC] at (475,91.5) {Which choice?};
\fgSq{458}{103.5}{\fgB}\node[fgS] at (475,102) {or};\fgSq{485}{103.5}{\fgA}
\path[draw=fgArrow!80,dash pattern=on 2pt off 1.4pt,line width=0.5pt,rounded corners=3pt] (374,124) rectangle (510,178);
\node[fgT] at (442,141) {Open-source check};
\node[fgS] at (442,154) {Local vs.\ sibling precedent};
\node[fgS] at (442,165) {in real REST API repositories};

\draw[fgline] (129.7,92.5) -- (138,92.5);
\draw[fgline,rounded corners=2pt] (153.7,39.5) -- (138,39.5) -- (138,155.5) -- (153.7,155.5);
\foreach \y in {39.5,92.5,155.5}{\draw[fgmain] (147,\y) -- (153.7,\y);}
\draw[fgline,rounded corners=2pt] (346,155.5) -- (360,155.5) -- (360,39.5);
\draw[fgline] (346,92.5) -- (360,92.5);
\draw[fgmain] (346,39.5) -- (373.7,39.5);
\end{tikzpicture}}
\caption{Research approach. Each task admits two functionally acceptable implementations (baseline in blue, alternative in orange). Within the controlled study, Stage~1 measures default choices without relevant examples, Stage~2 varies repository precedents in local and sibling-resource files (including which file holds the example at a fixed count), and Stage~3 adds task guidance that contradicts existing code. Every edit is scored separately for functional success and implementation choice; the local-versus-sibling comparison is also checked in open-source repositories.}
\label{fig:overview_new}
\end{figure*}

\subsection{Overview}
\label{sec:approach-overview}
We design a three-stage controlled study that separates the sources of guidance a coding agent may follow when it extends an existing repository (Figure~\ref{fig:overview_new}). The design rests on one principle: every stage keeps the functional task and its hidden tests unchanged and varies only the guidance available to the agent. A change in the agent's implementation choice can therefore be attributed to the varied guidance rather than to a change in what the code must do.

\vspace{.3em}
\noindent{\bf Study Object.}
Each study task admits two functionally acceptable implementations, which we call an \emph{implementation pair}. Before running any agent, we designate one implementation as the \emph{baseline} and the other as the \emph{alternative}. These labels are fixed in advance and do not encode which implementation an agent prefers. Pair-specific hidden tests accept either implementation, so the functional requirements stay fixed while the implementation choice remains open. For example, in the ID serialization case, the tests verify that the returned identifier matches the expected task ID, accepting either its numeric or string representation. We study six pairs spanning naming, data representation, and code structure (Table~\ref{tab:pairs}), and use ID serialization from Section~\ref{sec:motivation} as the running example: a new endpoint may serialize task identifiers as integers (baseline) or strings (alternative).

\begin{table}[t]
\centering
\footnotesize
\definecolor{fgMuted}{HTML}{555D69}%
\definecolor{fgBase}{HTML}{4A72A8}\definecolor{fgBaseFill}{HTML}{DCE5F1}\definecolor{fgBaseText}{HTML}{2F5486}%
\definecolor{fgAlt}{HTML}{EE9A45}\definecolor{fgAltFill}{HTML}{FBE6D0}\definecolor{fgAltText}{HTML}{9A5512}%
\def\fgChip#1#2#3{\tikz[baseline=(c.base)]\node(c)[fill=#1,text=#2,rounded corners=1.6pt,
  inner xsep=2.6pt,inner ysep=1.8pt,font=\ttfamily\scriptsize]{\strut#3};}%
\def\B#1{\fgChip{fgBaseFill}{fgBaseText}{#1}}\def\A#1{\fgChip{fgAltFill}{fgAltText}{#1}}%
\def\fgKey#1{\tikz\fill[#1,rounded corners=0.6pt] (0,-0.5pt) rectangle (6pt,5.5pt);}%
\def\fgCat#1{\multirow{2}{*}{\itshape\color{fgMuted}#1}}%
\setlength{\tabcolsep}{7pt}%
\renewcommand{\arraystretch}{1.3}%
\caption{Implementation pairs. Both implementations in a pair pass the same hidden functional tests; adoption refers to the alternative.}
\label{tab:pairs}
\begin{tabular}{@{}llll@{}}
\toprule
\textbf{Category} & \textbf{Pair} & \fgKey{fgBase}\,\textbf{Baseline} & \fgKey{fgAlt}\,\textbf{Alternative} \\
\midrule
\fgCat{Naming}              & Field casing      & \B{project\_id}      & \A{projectId} \\
                            & Boolean naming    & \B{completed}        & \A{is\_completed} \\
\addlinespace[3pt]
\fgCat{Data representation} & ID serialization  & \B{"id": 1}          & \A{"id": "1"} \\
                            & Response envelope & \B{\{...\}}          & \A{\{"data": \{...\}\}} \\
\addlinespace[3pt]
\fgCat{Code structure}      & Error handling    & \B{return ..., 404}  & \A{raise NotFound} \\
                            & Dispatch          & \B{if/elif chain}    & \A{LABELS[priority]} \\
\bottomrule
\end{tabular}
\par\vspace{3pt}
\end{table}

\vspace{.3em}
\noindent{\bf Controlled Repository.}
All three stages use a purpose-built Flask Task Manager API. In its full form, the repository contains three endpoints for \codefontsmall{tasks} (create, update, and list) and three for \codefontsmall{projects} (create, update, and list), each in its own file, together with the data model, database setup, and application scaffold. The agent is asked to add a new endpoint for \codefontsmall{task}, \texttt{GET /projects/\textless id\textgreater/tasks/\textless id\textgreater}. We therefore call the existing \codefontsmall{task} endpoints \emph{local} files, because they belong to the resource being extended, and the \codefontsmall{project} endpoints \emph{sibling-resource} files, because they implement an analogous resource that the new endpoint does not modify. Where an implementation pair requires additional functionality, we minimally extend the agent’s assignment. For example, the field-casing case additionally requires the endpoint to return the parent \codefontsmall{project} ID, and the Boolean-naming case requires adding a domain-model field.

\vspace{.3em}
\noindent{\bf Three Stages.}
Each stage (Figure~\ref{fig:overview_new}) isolates one source of guidance. Stage~1 removes all existing endpoints to measure default choices (Section~\ref{sec:approach-default}); Stage~2 controls how many and which local and sibling-resource files use the alternative (Section~\ref{sec:approach-precedent}); and Stage~3 sets every endpoint opposite a target and supplies explicit guidance through the task description or the harness's native instruction file (Section~\ref{sec:approach-prompt}). All stages assess functional success and implementation choice, as described in Section~\ref{sec:approach-measure}.

\subsection{Stage 1: Establishing Default Preferences}
\label{sec:approach-default}

\noindent{\bf Zero-Precedent Repository.}
Stage~1 establishes agents' default implementation choices without relevant repository precedents.
We remove all six existing task and project endpoint files from the controlled repository while retaining the data model, database setup, and application scaffold.
The resulting \emph{zero-precedent} repository remains sufficient for implementing the new endpoint but contains no relevant endpoint examples. We construct one variant per implementation pair, retaining the necessary pair-specific task extensions.

For each pair and agent, we run independent sessions on fresh repository copies, using the same functional task as in Stage~2 without prescribing an implementation choice. We classify the resulting implementations (Section~\ref{sec:approach-measure}) and define the option chosen in more than half of an agent's sessions as its default. These defaults provide a reference for measuring the influence of repository precedent and explicit guidance in subsequent stages, independently of the predefined baseline and alternative labels.

In the running example, the \texttt{task} model retains its integer primary key, but no existing handler demonstrates whether IDs should be returned as \texttt{task.id} or \texttt{str(task.id)}. In this setting, all four models default to returning numeric IDs (\texttt{task.id}), as confirmed by the zero-precedent results in Section~\ref{sec:rq1}.

\subsection{Stage 2: Manipulating Repository Precedents}
\label{sec:approach-precedent}

Stage~2 restores the six endpoints and rewrites the construct under study in a controlled subset of them. For ID serialization, we change an existing endpoint from \texttt{"id": task.id} to \texttt{"id": str(task.id)}; all other code in that endpoint stays identical. The other pairs are rewritten analogously, changing only the construct under study. Throughout Stage~2, the task description asks the agent to follow the conventions of the existing code without naming the implementation choice, mirroring common contribution guidelines. Any shift toward the alternative must therefore come from the code itself.

\vspace{.3em}
\noindent{\bf Support Level and Conditions.}
We call the number of files in a group (local or sibling-resource) that use the alternative the group's \emph{support level} ($k/3$). For each pair, we construct eight conditions (Figure~\ref{fig:precedent-interventions}, upper rows):
\begin{itemize}
    \item \emph{Agreeing groups:} all six existing endpoints use the baseline (“Both baseline”; $0/3$, $0/3$) or the alternative (“Both alternative”; $3/3$, $3/3$).
    \item \emph{Local support:} local support $1/3$, $2/3$, or $3/3$, with all sibling-resource files at the baseline.
    \item \emph{Sibling-resource support:} sibling-resource support $1/3$, $2/3$, or $3/3$, with all local files at the baseline.
\end{itemize}
We call this design \emph{bidirectional} because each group is varied while the other is held at the baseline, so the influence of the two groups can be compared at the same support level. Whenever one group supports the alternative, the two groups conflict; at $3/3$ they disagree completely, and comparing local $3/3$ with sibling-resource $3/3$ shows which group prevails in a direct conflict. Comparing these conflicting conditions with the agreeing ones shows whether agreement between the groups strengthens adoption.

\vspace{.3em}
\noindent{\bf File Identity.}
In the eight conditions, each intermediate support level uses one fixed assignment of the alternative to files; at local support $1/3$, for example, it appears in \texttt{tasks\_list} (Figure~\ref{fig:precedent-interventions}, ``default''). Raising the support level therefore also changes \emph{which} files carry the alternative, so example count and file identity are confounded. To separate them, we hold the support level fixed and vary only the placement: at $1/3$ support, we place the alternative in the create, update, or list endpoint (Figure~\ref{fig:precedent-interventions}, lower rows); at $2/3$ support, in each of the three pairs of these endpoints. We first conduct this analysis exploratorily for ID serialization with one agent, including a control that changes only module registration and import order, and then replicate it in three further pair--agent settings (Section~\ref{sec:rq2}).

\begin{figure}[t]
\centering
\footnotesize
\definecolor{fgInk}{HTML}{1F2328}%
\definecolor{fgMuted}{HTML}{555D69}%
\definecolor{fgBase}{HTML}{4A72A8}%
\definecolor{fgAlt}{HTML}{EE9A45}%
\def\fgSqr#1{\tikz\fill[#1,rounded corners=0.7pt] (0,-0.7pt) rectangle (7pt,6.3pt);}%
\def\N{\fgSqr{fgBase}}\def\S{\fgSqr{fgAlt}}%
\def\fgHd#1{\makebox[24pt]{\scriptsize\color{fgMuted}#1}}%
\def\fgGrp#1{\multicolumn{7}{@{}l}{\scriptsize\itshape\color{fgMuted}#1}}%
\setlength{\tabcolsep}{5pt}%
\renewcommand{\arraystretch}{1.2}%
\begin{tabular}{@{}l @{\hspace{14pt}} ccc @{\hspace{20pt}} ccc@{}}
\toprule
 & \multicolumn{3}{c@{\hspace{20pt}}}{\textbf{Tasks} (local resource)} & \multicolumn{3}{c@{}}{\textbf{Projects} (sibling resource)} \\
\cmidrule(r{16pt}){2-4}\cmidrule(l{0pt}){5-7}
 & \fgHd{create} & \fgHd{update} & \fgHd{list} & \fgHd{create} & \fgHd{update} & \fgHd{list} \\
\midrule
\fgGrp{Varying precedent support} \\
All numeric          & \N & \N & \N & \N & \N & \N \\
Local support $1/3$  & \N & \N & \S & \N & \N & \N \\
Local support $2/3$  & \N & \S & \S & \N & \N & \N \\
Sibling support $1/3$& \N & \N & \N & \N & \N & \S \\
All string           & \S & \S & \S & \S & \S & \S \\
\midrule
\fgGrp{Fixed $1/3$ local support, different file} \\
In create              & \S & \N & \N & \N & \N & \N \\
In update              & \N & \S & \N & \N & \N & \N \\
In list (default)      & \N & \N & \S & \N & \N & \N \\
\bottomrule
\end{tabular}%
\par\vspace{6pt}
{\scriptsize\color{fgMuted}\N~numeric ID (\texttt{task.id})\qquad\S~string ID (\texttt{str(task.id)})}
\caption{Illustrative repository interventions for ID serialization. Each square is one existing endpoint file. The upper rows vary how many local or sibling-resource files use string IDs; the lower rows hold the local support level at $1/3$ and change only which file carries the alternative (``default'' marks the assignment used for local support $1/3$ above). The requested new endpoint is absent from every variant.}
\label{fig:precedent-interventions}
\end{figure}

\subsection{Stage 3: Introducing Explicit Task Guidance}
\label{sec:approach-prompt}

Stage~3 asks whether new task intent can override repository precedent when the two conflict. We designate a \emph{target} implementation and set all six existing endpoints to the opposite implementation, so that no file in the repository supports the target and any adoption must come from the guidance. In the running example, all existing endpoints serialize IDs as integers, whereas the target uses strings. The functional task and hidden tests remain unchanged.

We compare four guidance conditions, all supplied in the initial task description:
\begin{itemize}
    \item \textbf{C0 (no guidance):} the task does not mention the target; this is the control.
    \item \textbf{C1 (written rule):} a statement of the target and its priority over existing code, without code, such as \emph{``Serialize task identifiers as strings in the new endpoint. This instruction takes precedence over the ID representations used by existing handlers.''}
    \item \textbf{C2 (code example):} a snippet that demonstrates the target, such as \texttt{"id": str(task.id)}, without stating its priority.
    \item \textbf{C3 (rule and example):} both C1 and C2.
\end{itemize}
Comparing C1 with C2 separates \emph{stating} which implementation should govern a new change from \emph{demonstrating} what that implementation looks like; comparing C3 with C1 and C2 shows whether the two forms complement each other. A separate experiment varies the delivery channel: the same written rule appears either in the task description or in the harness's native instruction file (\texttt{CLAUDE.md} or \texttt{AGENTS.md}), with the task description otherwise unchanged.

\subsection{Measuring Outcomes}
\label{sec:approach-measure}

We evaluate each generated edit on two separate outcomes. \emph{Functional success} indicates whether the edit passes the pair-specific hidden tests, which accept either implementation; sessions that produce no implementation count as functional failures. \emph{Implementation choice} indicates which alternative the edit adopts, determined by pair-specific classifiers that inspect the produced code. Outputs that use neither or both alternatives are reported as unclassifiable rather than assigned a label. For Stages~1 and~2, we report the \emph{adoption rate}: the share of classifiable outputs that use the designated alternative. For Stage~3, where the task prescribes a target, we use a \emph{joint outcome} that requires both functional success and adoption of the target. Separating the two outcomes prevents a passing test suite from hiding which implementation an agent selected.

\subsection{Open-Source Repository Validation}
\label{sec:approach-validation}

The controlled repository is small and written in a single language. We therefore test whether its central Stage~2 contrast, local versus sibling-resource precedent, also holds in established codebases. We select four open-source REST API repositories in Python, Go, and TypeScript and study two implementation choices that occur across their frameworks. In \emph{error handling}, an endpoint either propagates a typed error to a central handler (P) or constructs the error response directly (D); for example, it may raise a not-found error that a central handler converts into a response, or build an HTTP 404 response itself. In \emph{response envelope}, an endpoint returns either a flat JSON object (F) or the payload under a sole top-level \texttt{data} key (D).

Following Stage~2, we rewrite the existing handlers of the resource being extended (local) and of an analogous resource (sibling) to an assigned pattern, independently of each other. %
This yields four configurations in which the two groups agree or conflict (e.g., PP, PD, DP, and DD, with the local pattern first); in a repository without a sibling resource, only the local pattern varies. The agent then implements a new local endpoint without guidance about the choice. We evaluate functional success with repository-specific hidden tests and classify the resulting implementation. Because each repository fixes the number and identity of its handlers, this study tests the influence of local versus sibling-resource precedent, but not the support level or file placement, for which the controlled repository remains the source of evidence. 
We also test a Stage~3 counterpart in the same repositories. With every local handler returning a flat payload, we ask whether a rule in the harness's native instruction file makes the agent wrap the payload of a new endpoint under a sole top-level \texttt{data} key, compared with a contemporaneous no-guidance control. Because each repository fixes its own handlers, this test measures the effect of the rule against real conflicting code, not the support level or file placement.

\section{Evaluation}\label{sec:evaluation}
\newcommand{\implication}{\par\smallskip\noindent\textit{Implication:}\ }
\newcommand{\runhead}[1]{\par\vspace{.3em}\noindent\textbf{#1}\ }

We answer four research questions:

\begin{itemize}
    \item \textbf{RQ1 (Default Preferences):} Which implementation do coding agents choose among functionally equivalent alternatives without relevant repository precedent?
    \item \textbf{RQ2 (Local Precedent):} How do implementation patterns in task-related code influence agents' choices?
    \item \textbf{RQ3 (Sibling-Resource Precedent):} How do patterns in sibling resources affect implementation choices, particularly when they conflict with local precedent?
    \item \textbf{RQ4 (Explicit Guidance):} Can explicit rules or examples steer agents toward a specified implementation despite conflicting repository precedent?
\end{itemize}
\noindent RQ1--RQ4 use the controlled repository; Section~\ref{sec:osr} tests the local-versus-sibling contrast of RQ2 and RQ3, and the effect of an explicit rule (RQ4), in open-source repositories.

\begin{table}[t]
\centering
\caption{Experimental scale: 5,625 trials in 375 cells, each run in 15 independent sessions on fresh repository instances.}
\label{tab:scale}
\small
\setlength{\tabcolsep}{6pt}
\begin{tabular*}{0.85\textwidth}{@{}l@{\extracolsep{\fill}}lr@{}}
\toprule
Study & Design & Trials \\
\midrule
\multicolumn{3}{@{}l}{\textit{Controlled repository}} \\
\hspace{0.6em}RQ1: zero precedent & 6 pairs $\times$ 4 agents                    & 360 \\
\hspace{0.6em}RQ2/3: support levels & 6 pairs $\times$ 4 agents $\times$ 8 cond. & 2,880 \\
\hspace{0.6em}RQ2: file identity (expl.) & 4 placements + order control                 & 75 \\
\hspace{0.6em}RQ2: file identity (repl.) & 3 settings $\times$ 6 placements             & 270 \\
\hspace{0.6em}RQ4: explicit guidance & 2 pairs $\times$ 2 agents $\times$ 4 cond.   & 240 \\
\hspace{0.6em}RQ4: native-file rule & 2 pairs $\times$ 2 agents $\times$ 3 cond.   & 180 \\
\addlinespace
\multicolumn{3}{@{}l}{\textit{Open-source repositories}} \\
\hspace{0.6em}Error handling            & 4 repos $\times$ 4 assign. $\times$ 4 agents & 960 \\
\hspace{0.6em}Response envelope         & 14 assign.\textsuperscript{a} $\times$ 2 agents & 420 \\
\hspace{0.6em}Native-file rule (envelope)  & 4 repos $\times$ 2 agents $\times$ 2 cond.  & 240 \\
\midrule
\textbf{Total}                  &                                              & \textbf{5,625} \\
\bottomrule
\end{tabular*}
\\[2pt]
\parbox{0.85\textwidth}{\footnotesize \textsuperscript{a}4 assignments in each of 3 repositories with a sibling resource, 2 in \texttt{fastapi-clean}.}
\end{table}

\subsection{Experimental Setup}\label{sec:overview}\label{sec:study-fixture}

\runhead{Agents.}\label{sec:exp-validity}
We run Claude Sonnet-4-6 and Claude Opus-4-8 through Claude Code~\cite{anthropic-claude-code}, and \texttt{gpt-5.6-sol} and \texttt{gpt-6-sol} through Codex CLI~\cite{openai-codex-cli}. We refer to each model--harness combination as an \emph{agent}.

\runhead{Trials.}
Each trial is an independent agent session on a fresh repository instance: the agent explores and edits the repository autonomously, and we run the saved result against the hidden functional tests and the pair-specific classifier. Each experimental cell is run 15 times, for 5,625 trials in total (Table~\ref{tab:scale}). 

\runhead{Analysis.}\label{sec:study-analysis}
Adoption rates are computed over classifiable outputs. For RQ1--RQ3, we report exact binomial intervals, Cochran--Armitage tests for trends across support levels, and Fisher exact tests for pairwise comparisons; the tests specific to file placement, RQ4, and the open-source study are given in the corresponding subsections. Figure~\ref{fig:adoption} reports the adoption counts of every RQ1--RQ3 cell.

\definecolor{heatalt}{RGB}{217,95,2}   %
\providecommand{\heatcell}[4]{%
  \pgfmathtruncatemacro{\hpct}{round(100*#3/#4)}%
  \fill[heatalt!\hpct!white] (#1-0.5,#2-0.5) rectangle (#1+0.5,#2+0.5);%
  \draw[white,line width=0.6pt] (#1-0.5,#2-0.5) rectangle (#1+0.5,#2+0.5);%
  \ifnum\hpct>55\relax
    \node[text=white,font=\scriptsize\bfseries] at (#1,#2) {#3};%
  \else
    \node[text=black,font=\scriptsize] at (#1,#2) {#3};%
  \fi}
\providecommand{\heatblank}[2]{%
  \fill[black!8] (#1-0.5,#2-0.5) rectangle (#1+0.5,#2+0.5);%
  \draw[white,line width=0.6pt] (#1-0.5,#2-0.5) rectangle (#1+0.5,#2+0.5);%
  \node[text=black!50,font=\scriptsize] at (#1,#2) {--};}
\providecommand{\heatlegend}[3]{%
  \shade[left color=white,right color=heatalt] (#1,#2-0.3) rectangle (#1+#3,#2+0.3);%
  \draw[black!40,line width=0.3pt] (#1,#2-0.3) rectangle (#1+#3,#2+0.3);%
  \node[anchor=east,font=\scriptsize] at (#1-0.1,#2) {0/15};%
  \node[anchor=west,font=\scriptsize] at (#1+#3+0.1,#2) {15/15 adopt};}
\providecommand{\heatcellB}[4]{%
  \pgfmathtruncatemacro{\hpct}{round(100*#3/#4)}%
  \fill[heatalt!\hpct!white] (#1-0.5,#2-0.5) rectangle (#1+0.5,#2+0.5);%
  \draw[white,line width=0.8pt] (#1-0.5,#2-0.5) rectangle (#1+0.5,#2+0.5);%
  \ifnum\hpct>55\relax
    \node[text=white,font=\footnotesize\bfseries] at (#1,#2) {#3};%
  \else
    \node[text=black,font=\footnotesize] at (#1,#2) {#3};%
  \fi}
\providecommand{\heatlegendB}[3]{%
  \shade[left color=white,right color=heatalt] (#1,#2-0.2) rectangle (#1+#3,#2+0.2);%
  \draw[black!40,line width=0.3pt] (#1,#2-0.2) rectangle (#1+#3,#2+0.2);%
  \node[anchor=east,font=\scriptsize] at (#1-0.15,#2) {0/15};%
  \node[anchor=west,font=\scriptsize] at (#1+#3+0.15,#2) {15/15};}

\begin{figure*}[!t]
\centering
\resizebox{\textwidth}{!}{\begin{tikzpicture}[x=5.6mm,y=4.1mm]
\foreach \p [count=\g from 0] in {ID serialization,Field casing,Envelope,Boolean naming,Error handling,Dispatch}{
  \pgfmathsetmacro{\cx}{4*\g+1.5+0.4*\g}
  \node[font=\scriptsize\bfseries] at (\cx,1.85) {\p};
  \draw[black!50,line width=0.4pt] (\cx-1.9,1.35) -- (\cx+1.9,1.35);
  \foreach \m [count=\j from 0] in {S,O,G5,G6}{
    \pgfmathsetmacro{\mx}{4*\g+\j+0.4*\g}
    \node[font=\scriptsize] at (\mx,0.85) {\m};
  }
}
\foreach \y/\lab/\vals in {
  0/{Zero precedent}/{0,0,0,0, 0,0,0,0, 0,0,0,0, 0,0,0,0, 0,0,0,0, 14,15,15,15},
  1.35/{Both baseline}/{0,0,0,0, 0,0,0,0, 0,0,0,0, 0,0,0,0, 0,0,0,0, 0,0,0,0},
  2.35/{Local 1/3}/{10,2,0,1, 7,0,0,0, 0,0,0,0, 0,0,0,0, 8,0,8,11, 4,2,9,12},
  3.35/{Local 2/3}/{12,9,14,14, 14,3,7,12, 2,0,0,2, 1,0,0,0, 11,5,13,13, 8,5,15,15},
  4.35/{Local 3/3}/{15,15,15,15, 15,15,15,15, 8,9,15,15, 15,15,15,15, 15,15,15,15, 15,15,15,15},
  5.7/{Sibling 1/3}/{0,0,0,0, 0,0,0,0, 0,0,0,0, 0,0,0,0, 0,0,0,0, 0,0,0,0},
  6.7/{Sibling 2/3}/{0,0,0,0, 0,0,0,0, 0,0,0,0, 0,0,0,0, 0,0,0,0, 0,0,0,0},
  7.7/{Sibling 3/3}/{0,0,0,0, 0,0,0,0, 0,0,0,0, 0,0,0,0, 2,0,2,2, 4,0,0,3},
  9.05/{Both alternative}/{15,15,15,15, 15,15,15,15, 15,15,15,15, 14,15,15,15, 15,15,15,15, 15,15,15,15}%
}{
  \node[anchor=east,font=\footnotesize] at (-0.65,-\y) {\lab};
  \foreach \v [count=\c from 0] in \vals {
    \pgfmathsetmacro{\xx}{\c+0.4*int(\c/4)}
    \heatcell{\xx}{-\y}{\v}{15}
  }
}
\draw[black!40,line width=0.4pt] (-5.2,-0.675) -- (25.5,-0.675);
\draw[black!40,line width=0.4pt] (-5.2,-5.025) -- (25.5,-5.025);
\draw[black!40,line width=0.4pt] (-5.2,-8.375) -- (25.5,-8.375);
\node[font=\tiny] at (1+0.36,-2.35+0.25) {*};      %
\node[font=\tiny] at (20+2.0+0.36,-6.7+0.25) {*};  %
\heatlegend{1.2}{-10.4}{4}
\node[anchor=west,font=\scriptsize] at (9.5,-10.4) {S = Sonnet-4-6, O = Opus-4-8, G5 = \texttt{gpt-5.6-sol}, G6 = \texttt{gpt-6-sol}};
\end{tikzpicture}}
\caption{Adoption of the designated alternative in the controlled repository (RQ1--RQ3). Each cell gives the number of runs, out of 15 classifiable runs (* = 14), that adopt the alternative; shading is proportional to the adoption rate. ``Local $k/3$'' and ``Sibling $k/3$'' vary the number of files in that group that use the alternative while the other group stays at the baseline; at intermediate levels the alternative appears in a fixed set of files. For dispatch, the designated alternative (dictionary lookup) is the agents' default, so the zero-precedent row is shaded.}
\label{fig:adoption}
\end{figure*}

\subsection{RQ1: Default Implementation Preferences}\label{sec:rq1}

\runhead{Setup.}
We run each pair on its zero-precedent repository (Section~\ref{sec:approach-default}) with all four agents, for 360 trials. All 360 runs pass the hidden tests and are classifiable.

\runhead{Default Choices.}
Without repository precedent, agents serialize identifiers as integers, name response fields in \texttt{snake\_case}, return the resource as a flat JSON object, name Boolean fields without an \texttt{is\_} prefix, construct error responses directly in the handler, and dispatch priority labels through a dictionary lookup (Table~\ref{tab:defaults}). In dispatch, the default is the designated alternative, which is why the corresponding zero-precedent cells in Figure~\ref{fig:adoption} are shaded. The six defaults share one property: each is the most direct expression of the required behavior in Python and Flask. Agents return the model attribute as stored (\texttt{task.id}, \texttt{project\_id}), return the payload without a wrapper, use the shorter Boolean name, handle an error where it arises, and map values through a lookup table. Each alternative instead adds a conversion, a wrapper, a naming convention, or an indirection that the task itself does not motivate. Choosing such an alternative therefore requires information from outside the task, which is precisely what repository precedent and task guidance supply.

\runhead{Consistency.}
The defaults are near-deterministic and agent-independent. Of the 360 runs, 359 choose the default, and 23 of the 24 pair--agent cells do so unanimously; the only deviation is a single Sonnet run on dispatch. All four agents, from two providers and running in two harnesses, share the same default in every pair. Differences among agents in RQ2--RQ4 therefore do not stem from different starting preferences.

\runhead{Defaults vs. Precedent.}
Repository precedent overrides these defaults. When all six existing endpoints use the implementation opposite the default, agents follow the repository in 359 of 360 runs: 299/300 in the five pairs whose default is the baseline (``Both alternative'' in Figure~\ref{fig:adoption}) and 60/60 in dispatch (``Both baseline''). Dictionary dispatch, chosen in 59 of 60 zero-precedent runs, is chosen in none of 60 runs once every endpoint uses an \texttt{if}/\texttt{elif} chain.

\rqanswer{Answer to RQ1}{Without relevant repository precedents, all four agents favor the same implementation in each of the six pairs: integer IDs, \texttt{snake\_case} fields, flat responses, unprefixed Boolean names, errors returned directly in the handler, and dictionary-based dispatch. These default choices are nearly deterministic (359/360 trials). Yet when all existing endpoints use the opposite implementation, agents follow repository precedent in 359/360 trials.
\implication Implementation preferences measured without relevant repository precedents do not necessarily predict agents' choices when extending existing code.}

\begin{table}[!b]
\centering
\caption{RQ1: default implementation in the zero-precedent repository, and the number of runs (of 15) in which each agent chooses it. ``Label'' gives the predefined role of the default in each pair (B = baseline, A = alternative). S = Sonnet-4-6, O = Opus-4-8, G5 = gpt-5.6-sol, G6 = gpt-6-sol.}
\label{tab:defaults}
\small
\setlength{\tabcolsep}{4pt}
\renewcommand{\arraystretch}{0.95}
\begin{tabular*}{0.8\textwidth}{@{}l@{\extracolsep{\fill}}lcrrrr@{}}
\toprule
Pair & Default implementation & Label & S & O & G5 & G6 \\
\midrule
ID serialization  & integer: \texttt{"id": 1}          & B & 15 & 15 & 15 & 15 \\
Field casing      & \texttt{snake\_case}: \texttt{project\_id} & B & 15 & 15 & 15 & 15 \\
Envelope          & flat: \texttt{\{...\}}               & B & 15 & 15 & 15 & 15 \\
Boolean naming    & no prefix: \texttt{completed}       & B & 15 & 15 & 15 & 15 \\
Error handling    & direct: \texttt{return ..., 404}    & B & 15 & 15 & 15 & 15 \\
Dispatch          & lookup: \texttt{LABELS[p]}          & A & 14 & 15 & 15 & 15 \\
\bottomrule
\end{tabular*}
\end{table}

\subsection{RQ2: Influence of Local Precedent}\label{sec:rq2}

\runhead{Setup.}\label{sec:study-precedent}
We use the local-support conditions of Stage~2 (Section~\ref{sec:approach-precedent}), in which zero to three local files use the alternative while the sibling-resource files stay at the baseline, for all six pairs and four agents (2,880 trials shared with RQ3). For file placement, we move the alternative among local files at $1/3$ and $2/3$ support, first exploratorily for ID serialization with Sonnet (75 trials, including an import-order control) and then in three replication settings, ID serialization with \texttt{gpt-6-sol} and with Opus-4-8, and field casing with Sonnet (270 trials). We compare placements in the replications with omnibus within-block Monte Carlo randomization tests (99,999 randomizations) and Holm correction over the six registered tests (two support levels in each setting).

\runhead{Support Level.}
Adoption increases with local support. Pooled over six pairs and four agents, adoption of the alternative rises from 0\% (0/360) without local support to 21\% (74/359) at $1/3$, 49\% (175/360) at $2/3$, and 96\% (347/360) at $3/3$ (Figure~\ref{fig:adoption}). Adoption never decreases with local support in any of the 24 pair--agent combinations, and Cochran--Armitage tests find positive trends for all six pairs with Sonnet ($p\le5.6\times10^{-5}$) and Opus ($p\le1.3\times10^{-5}$). The curves share a direction but not a threshold. A single local file suffices for Sonnet to adopt string IDs (10/15), centrally handled errors (8/15), and \texttt{camelCase} (7/15), whereas Boolean naming stays at or below 1/15 through $2/3$ support for every agent and reaches 15/15 only at $3/3$. Response envelope never reaches full adoption for the Claude agents (Sonnet 8/15, Opus 9/15 at $3/3$). Agents also differ in which pairs they respond to: at $1/3$ support, Sonnet adopts string IDs more often than \texttt{gpt-6-sol} (10/15 vs.\ 1/15, Fisher exact $p=0.0017$), whereas for dictionary dispatch the order reverses (4/15 vs.\ 12/15, $p=0.0092$).\label{sec:models}

\runhead{File Placement.}\label{sec:file-identity}
Holding local support fixed, changing which files contain supporting examples substantially changes adoption in three of four tested pair--agent settings (Figure~\ref{fig:fileid}). In the exploratory ID-serialization experiment with Sonnet, a single supporting file yields 1/15--10/15 adoptions depending on placement ($p=9.6\times10^{-4}$). Two of three replications confirm the effect: with \texttt{gpt-6-sol}, relocating two ID-serialization examples changes adoption from 0/15 to 15/15 (Holm-adjusted $p\le6\times10^{-5}$); with Sonnet, relocating one field-casing example changes adoption from 0/15 to 11/15 ($p\le6\times10^{-5}$). The Opus ID-serialization replication shows no detectable placement effect ($p=0.148$).

Placement can even outweigh example count. For field casing with Sonnet, one example in \texttt{tasks\_list} yields 11/15 adoptions, whereas two examples in the other local files yield none. Where placement matters, the most influential configurations include \texttt{tasks\_list}, although agents differ in how much supporting evidence they require. These results show that example frequency alone cannot explain the influence of repository precedent.

\rqanswer{Answer to RQ2}{Local precedent strongly influences implementation choices: adoption rises from 0\% without local support to 96\% with full local support, with no decrease across support levels in any of the 24 pair--agent settings. However, at a fixed support level, changing file placement can shift adoption from 0\% to 100\%, with detectable effects in three of four tested settings.
\implication Example frequency alone is insufficient to characterize the influence of repository precedent. Where examples appear can matter as much as, or even more than, how many appear.}

\subsection{RQ3: Influence of Sibling-Resource Precedent}\label{sec:rq3}

\runhead{Setup.}
We use the sibling-resource conditions of Stage~2, in which one to three sibling-resource files use the alternative while all local files stay at the baseline (1,080 trials), and compare them with the local-support and agreeing conditions of RQ2.

\definecolor{heatalt}{RGB}{217,95,2}   %
\providecommand{\heatcell}[4]{%
  \pgfmathtruncatemacro{\hpct}{round(100*#3/#4)}%
  \fill[heatalt!\hpct!white] (#1-0.5,#2-0.5) rectangle (#1+0.5,#2+0.5);%
  \draw[white,line width=0.6pt] (#1-0.5,#2-0.5) rectangle (#1+0.5,#2+0.5);%
  \ifnum\hpct>55\relax
    \node[text=white,font=\scriptsize\bfseries] at (#1,#2) {#3};%
  \else
    \node[text=black,font=\scriptsize] at (#1,#2) {#3};%
  \fi}
\providecommand{\heatblank}[2]{%
  \fill[black!8] (#1-0.5,#2-0.5) rectangle (#1+0.5,#2+0.5);%
  \draw[white,line width=0.6pt] (#1-0.5,#2-0.5) rectangle (#1+0.5,#2+0.5);%
  \node[text=black!50,font=\scriptsize] at (#1,#2) {--};}
\providecommand{\heatlegend}[3]{%
  \shade[left color=white,right color=heatalt] (#1,#2-0.3) rectangle (#1+#3,#2+0.3);%
  \draw[black!40,line width=0.3pt] (#1,#2-0.3) rectangle (#1+#3,#2+0.3);%
  \node[anchor=east,font=\scriptsize] at (#1-0.1,#2) {0/15};%
  \node[anchor=west,font=\scriptsize] at (#1+#3+0.1,#2) {15/15 adopt};}
\providecommand{\heatcellB}[4]{%
  \pgfmathtruncatemacro{\hpct}{round(100*#3/#4)}%
  \fill[heatalt!\hpct!white] (#1-0.5,#2-0.5) rectangle (#1+0.5,#2+0.5);%
  \draw[white,line width=0.8pt] (#1-0.5,#2-0.5) rectangle (#1+0.5,#2+0.5);%
  \ifnum\hpct>55\relax
    \node[text=white,font=\footnotesize\bfseries] at (#1,#2) {#3};%
  \else
    \node[text=black,font=\footnotesize] at (#1,#2) {#3};%
  \fi}
\providecommand{\heatlegendB}[3]{%
  \shade[left color=white,right color=heatalt] (#1,#2-0.2) rectangle (#1+#3,#2+0.2);%
  \draw[black!40,line width=0.3pt] (#1,#2-0.2) rectangle (#1+#3,#2+0.2);%
  \node[anchor=east,font=\scriptsize] at (#1-0.15,#2) {0/15};%
  \node[anchor=west,font=\scriptsize] at (#1+#3+0.15,#2) {15/15};}

\begin{figure}[t]
\centering
\begin{tikzpicture}[x=9.5mm,y=5mm]
\node[font=\footnotesize\itshape] at (1,2.5) {One file (1/3)};
\node[font=\footnotesize\itshape] at (6.4,2.5) {Two files (2/3)};
\foreach \h/\xx in {C/0,U/1,\textbf{L}/2,C+U/5.4,\textbf{C+L}/6.4,\textbf{U+L}/7.4}{
  \node[font=\footnotesize] at (\xx,1.2) {\h}; }
\node[font=\footnotesize\itshape,text=black!60,anchor=west] at (2.65,1.2) {$p$};
\node[font=\footnotesize\itshape,text=black!60,anchor=west] at (8.05,1.2) {$p$};
\foreach \y/\lab/\vals/\pa/\rvals/\pb in {
  0/{ID / Sonnet-4-6 (expl.)}/{1,3,10}/{$9.6\times10^{-4}$}/{}/{$0.11$},
  1/{ID / \texttt{gpt-6-sol}}/{0,0,0}/{$1$}/{0,8,15}/{$\le6\times10^{-5}$},
  2/{ID / Opus-4-8}/{0,0,0}/{$1$}/{14,11,15}/{$0.148$},
  3/{Casing / Sonnet-4-6}/{0,0,11}/{$\le6\times10^{-5}$}/{0,11,9}/{$3.3\times10^{-4}$}%
}{
  \node[anchor=east,font=\footnotesize] at (-0.65,-\y) {\lab};
  \foreach \v [count=\c from 0] in \vals { \heatcellB{\c}{-\y}{\v}{15} }
  \node[anchor=west,font=\scriptsize,text=black!60] at (2.65,-\y) {\pa};
  \node[anchor=west,font=\scriptsize,text=black!60] at (8.05,-\y) {\pb};
}
\foreach \y/\vals in {1/{0,8,15}, 2/{14,11,15}, 3/{0,11,9}}{
  \foreach \v [count=\c from 0] in \vals { \pgfmathsetmacro{\xx}{5.4+\c}\heatcellB{\xx}{-\y}{\v}{15} }
}
\foreach \c in {0,1,2}{ \pgfmathsetmacro{\xx}{5.4+\c}\node[font=\footnotesize,text=black!50] at (\xx,0) {--}; }
\end{tikzpicture}
\caption{File-placement effects at fixed local support (15 trials per cell). C, U, L denote \texttt{tasks\_create}, \texttt{tasks\_update}, \texttt{tasks\_list}; bold headers indicate placements containing L. Rows show one exploration and three replications; the $p$ column gives the placement test within each support level ($\chi^2$ for exploration, Holm-adjusted within-block randomization for replications). ``--'' denotes omitted results (11--15/15).}
\label{fig:fileid}
\end{figure}

\runhead{Conflicting Sibling Precedent.}
Sibling-resource precedent rarely overrides local code. Across 1,079 classifiable sibling-only trials, agents adopt the sibling-supported alternative 13 times (1.2\%): 6/269 for Sonnet, 5/270 for \texttt{gpt-6-sol}, 2/270 for \texttt{gpt-5.6-sol}, and 0/270 for Opus. In four pairs (ID serialization, field casing, response envelope, and Boolean naming), each agent records 0/180 adoptions. All 13 adoptions occur at full sibling support, for centrally handled errors (2/15 each for Sonnet and both GPT agents) and dictionary dispatch (4/15 for Sonnet, 3/15 for \texttt{gpt-6-sol}). The location of the examples, not their number, decides the outcome: three supporting sibling-resource files yield 13/360 adoptions, whereas three supporting local files yield 347/360. Agents do see the sibling code. In all 45 Sonnet ID-serialization trials with sibling support, the \texttt{projects\_*} files appeared in context visible to the orchestrating agent, and 22 transcripts explicitly discuss the conflicting ID format; the agents nevertheless follow the local files.

\runhead{Agreeing Sibling Precedent.}
When sibling-resource code agrees with local code, it reinforces the local pattern. For response envelope, Sonnet adopts the alternative in 8/15 runs when only the local files use it and in 15/15 runs when both groups do (Fisher exact $p=0.0063$); Opus rises from 9/15 to 15/15. In the other pairs, local support alone already yields full or near-full adoption.

\rqanswer{Answer to RQ3}{When sibling-resource precedent conflicts with local code, agents adopt the sibling pattern in only 13/1,079 trials (1.2\%). With three supporting files, adoption reaches 347/360 for local precedents but only 13/360 for sibling-resource precedents. However, sibling-resource precedent can reinforce matching local patterns, increasing response-envelope adoption from 8/15 to 15/15 in one tested setting.
\implication Repository conventions do not necessarily propagate across resource boundaries through examples alone. Agents predominantly follow the implementation patterns of the resource they extend.}

\subsection{RQ4: Redirecting Choices with Explicit Guidance}\label{sec:rq4}
\runhead{Setup.} We test ID serialization and response envelope with Sonnet-4-6 and \texttt{gpt-6-sol}, with all six endpoints opposite the target. The content experiment runs 15 randomized blocks of C0--C3 per setting (240 trials); the delivery experiment compares no rule, a task-description rule, and a native-file rule (\texttt{CLAUDE.md}/\texttt{AGENTS.md}) (180 trials). Both use the joint outcome and exact block-conditional randomization tests with Holm correction.

\runhead{Rules vs. Examples.}
Without guidance (C0), no run adopts the target in any setting (0/60; Figure~\ref{fig:rq4}). A written rule that states the target and its priority over existing code (C1) yields 59 of 60 joint successes, 14--15/15 in every setting (Holm-adjusted $p\le0.00159$ vs.\ C0). The rule combined with an example (C3) yields 15/15 in every setting (risk difference 1.00; Holm-adjusted $p=2.44\times10^{-4}$ vs.\ C0), and adding the example yields no gain over the rule alone (C3 vs.\ C1, $p_{\mathrm{Holm}}=1$). An example alone (C2) is unreliable: it yields 38 of 60 joint successes, 15/15 for \texttt{gpt-6-sol} on both pairs but only 8/15 for Sonnet on response envelope and 0/15 on string IDs, the pair on which Sonnet followed local examples most readily in RQ2. An example thus acts as one more precedent that competes with the repository, whereas a stated priority resolves the conflict. Guidance does not cost functional success: 238 of the 240 runs pass the hidden tests.

\definecolor{heatalt}{RGB}{217,95,2}   %
\providecommand{\heatcell}[4]{%
  \pgfmathtruncatemacro{\hpct}{round(100*#3/#4)}%
  \fill[heatalt!\hpct!white] (#1-0.5,#2-0.5) rectangle (#1+0.5,#2+0.5);%
  \draw[white,line width=0.6pt] (#1-0.5,#2-0.5) rectangle (#1+0.5,#2+0.5);%
  \ifnum\hpct>55\relax
    \node[text=white,font=\scriptsize\bfseries] at (#1,#2) {#3};%
  \else
    \node[text=black,font=\scriptsize] at (#1,#2) {#3};%
  \fi}
\providecommand{\heatblank}[2]{%
  \fill[black!8] (#1-0.5,#2-0.5) rectangle (#1+0.5,#2+0.5);%
  \draw[white,line width=0.6pt] (#1-0.5,#2-0.5) rectangle (#1+0.5,#2+0.5);%
  \node[text=black!50,font=\scriptsize] at (#1,#2) {--};}
\providecommand{\heatlegend}[3]{%
  \shade[left color=white,right color=heatalt] (#1,#2-0.3) rectangle (#1+#3,#2+0.3);%
  \draw[black!40,line width=0.3pt] (#1,#2-0.3) rectangle (#1+#3,#2+0.3);%
  \node[anchor=east,font=\scriptsize] at (#1-0.1,#2) {0/15};%
  \node[anchor=west,font=\scriptsize] at (#1+#3+0.1,#2) {15/15 adopt};}
\providecommand{\heatcellB}[4]{%
  \pgfmathtruncatemacro{\hpct}{round(100*#3/#4)}%
  \fill[heatalt!\hpct!white] (#1-0.5,#2-0.5) rectangle (#1+0.5,#2+0.5);%
  \draw[white,line width=0.8pt] (#1-0.5,#2-0.5) rectangle (#1+0.5,#2+0.5);%
  \ifnum\hpct>55\relax
    \node[text=white,font=\footnotesize\bfseries] at (#1,#2) {#3};%
  \else
    \node[text=black,font=\footnotesize] at (#1,#2) {#3};%
  \fi}
\providecommand{\heatlegendB}[3]{%
  \shade[left color=white,right color=heatalt] (#1,#2-0.2) rectangle (#1+#3,#2+0.2);%
  \draw[black!40,line width=0.3pt] (#1,#2-0.2) rectangle (#1+#3,#2+0.2);%
  \node[anchor=east,font=\scriptsize] at (#1-0.15,#2) {0/15};%
  \node[anchor=west,font=\scriptsize] at (#1+#3+0.15,#2) {15/15};}

\begin{figure}[t]
\centering
\begin{tikzpicture}[x=10mm,y=5mm]
\node[font=\footnotesize\itshape] at (1.5,2.6) {Guidance content};
\node[font=\footnotesize\itshape] at (5.9,2.6) {Rule delivery};
\foreach \h/\d [count=\c from 0] in {C0/none,C1/rule,C2/example,C3/both}{
  \node[font=\footnotesize,align=center] at (\c,1.3) {\textbf{\h}\\\scriptsize\d};
}
\foreach \h/\xx in {none/4.9,task/5.9,file/6.9}{
  \node[font=\footnotesize,align=center] at (\xx,1.3) {\textbf{\h}\\\scriptsize\phantom{x}};
}
\foreach \y/\lab/\vals/\rvals in {
  0/{ID / Sonnet-4-6}/{0,15,0,15}/{0,15,15},
  1/{ID / \texttt{gpt-6-sol}}/{0,14,15,15}/{0,15,15},
  2/{Envelope / Sonnet-4-6}/{0,15,8,15}/{0,15,15},
  3/{Envelope / \texttt{gpt-6-sol}}/{0,15,15,15}/{0,15,15}%
}{
  \node[anchor=east,font=\footnotesize] at (-0.65,-\y) {\lab};
  \foreach \v [count=\c from 0] in \vals { \heatcellB{\c}{-\y}{\v}{15} }
  \foreach \v [count=\c from 0] in \rvals { \pgfmathsetmacro{\xx}{4.9+\c}\heatcellB{\xx}{-\y}{\v}{15} }
}
\draw[black!30,line width=0.4pt] (-0.5,-3.55) -- (3.5,-3.55);
\draw[black!30,line width=0.4pt] (4.4,-3.55) -- (7.4,-3.55);
\node[anchor=east,font=\footnotesize\bfseries] at (-0.65,-4.1) {Pooled (of 60)};
\foreach \v [count=\c from 0] in {0,59,38,60}{ \node[font=\footnotesize\bfseries] at (\c,-4.1) {\v}; }
\foreach \v [count=\c from 0] in {0,60,60}{ \pgfmathsetmacro{\xx}{4.9+\c}\node[font=\footnotesize\bfseries] at (\xx,-4.1) {\v}; }
\end{tikzpicture}
\caption{RQ4 joint success (passing functional tests and adopting the target) out of 15 trials per cell, with all six existing endpoints favoring the opposite implementation. Left (240 trials): C0 = no guidance, C1 = written rule, C2 = code example, C3 = both. Right (180 trials): no rule, task-description rule, or harness-loaded native-file rule (\texttt{CLAUDE.md}/\texttt{AGENTS.md}).}
\label{fig:rq4}
\end{figure}

\runhead{Rule Delivery.}
A rule delivered through the native instruction file is followed as reliably as a rule in the task description (Figure~\ref{fig:rq4}, right panel). Without a rule, no run succeeds (0/60); with the rule in the task description, all 60 runs do, and with the rule in the native instruction file, again all 60, 15/15 in every setting. The primary native-file versus no-rule contrasts each have a risk difference of 1.00 (Holm-adjusted $p=0.000244$).
The two channels do not differ detectably, but with every cell at the ceiling the data cannot show equivalence: the pooled 60/60 versus 60/60 comparison is compatible with a difference of up to about 6 percentage points in either direction (Newcombe interval). The task-rule condition also reproduces the earlier written-rule result (C1: 14--15/15 in each setting). 

\rqanswer{Answer to RQ4}{Despite unanimous conflicting repository precedent, written priority rules achieve both functional success and target adoption in 59/60 trials, compared with 38/60 for code examples alone and 0/60 without guidance. In a separate experiment, rules in the task description and harness-loaded native instruction files each achieve 60/60. In four open-source repositories, a native-file rule raises joint success from 0/120 to 117/120.
\implication Explicitly stating the priority of new task intent is more reliable than providing examples alone. Because both implementation choices satisfy the same functional requirements, verifying that intent requires measuring implementation choice alongside functional success.}

\subsection{Generalization to Open-Source Repositories}\label{sec:osr}

\runhead{Setup.}\label{sec:study-robustness}
We evaluate four REST API repositories spanning Python, Go, and TypeScript: FastAPI RealWorld~\cite{repo-fastapi-realworld}, FastAPI Clean Example~\cite{repo-fastapi-clean}, Go REST API~\cite{repo-go-rest}, and ExpressJS API~\cite{repo-expressjs-api} (Section~\ref{sec:approach-validation}). We apply the design of Section III-F to these four repositories. For error handling, all four agents run 15 trials for each of the four
local/sibling assignments (PP, PD, DP, DD) in each repository (960 trials).
For response envelope, Sonnet-4-6 and \texttt{gpt-6-sol} run the same design
(420 trials; \texttt{fastapi-clean} has no sibling resource, so only FN and DN).
For explicit guidance, a native-file rule is compared with a no-guidance
control in the same repositories and agents (240 trials). For statistical analysis, we use Wilson intervals and stratified Fisher and Mantel--Haenszel tests for error handling. Response-envelope comparisons use Holm-adjusted tests for local and sibling effects, with block-bootstrap intervals (99,999 draws). Explicit-guidance comparisons use paired exact tests with Holm correction across eight repository--agent settings.

\begin{table}[!b]
\begin{minipage}[t]{0.46\textwidth}\centering
\caption{Open-source error-handling study. Task success uses all 960 trials; local adoption uses 945 functionally successful, classifiable implementations. Agent and repository rows summarize the same trials.}
\label{tab:osr-qualified}
\footnotesize
\setlength{\tabcolsep}{3pt}
\resizebox{\linewidth}{!}{\begin{tabular}{lrrr}
\toprule
Group & Task success & Local adoption & Unclass. \\
\midrule
Sonnet-4-6 & 228/240 & 228/228 & 0 \\
Opus-4-8 & 240/240 & 240/240 & 0 \\
gpt-5.6-sol & 240/240 & 240/240 & 0 \\
gpt-6-sol & 240/240 & 237/237 & 3 \\
\midrule
FastAPI RealWorld & 240/240 & 240/240 & 0 \\
FastAPI Clean & 236/240 & 233/233 & 3 \\
Go REST & 233/240 & 233/233 & 0 \\
ExpressJS API & 239/240 & 239/239 & 0 \\
\midrule
\textbf{All} & \textbf{948/960} & \textbf{945/945} & \textbf{3} \\
\bottomrule
\end{tabular}}
\end{minipage}\hfill
\begin{minipage}[t]{0.5\textwidth}\centering
\caption{Open-source explicit guidance: runs (of 15) that both pass the hidden tests and wrap the payload under \texttt{data}, when all local handlers return flat payloads. No rule: contemporaneous control; native rule: the rule in \texttt{CLAUDE.md} (Sonnet-4-6) or \texttt{AGENTS.md} (\texttt{gpt-6-sol}).}
\label{tab:osr-guidance}
\footnotesize
\setlength{\tabcolsep}{4pt}
\resizebox{\linewidth}{!}{\begin{tabular}{lrrrr}
\toprule
 & \multicolumn{2}{c}{Sonnet-4-6} & \multicolumn{2}{c}{\texttt{gpt-6-sol}} \\
\cmidrule(lr){2-3}\cmidrule(lr){4-5}
Repository & No rule & Native rule & No rule & Native rule \\
\midrule
FastAPI RealWorld & 0/15 & 12/15 & 0/15 & 15/15 \\
Go REST & 0/15 & 15/15 & 0/15 & 15/15 \\
ExpressJS API & 0/15 & 15/15 & 0/15 & 15/15 \\
FastAPI Clean & 0/15 & 15/15 & 0/15 & 15/15 \\
\midrule
\textbf{Total} & 0/60 & 57/60 & 0/60 & 60/60 \\
\bottomrule
\end{tabular}}
\end{minipage}
\end{table}

\begin{table}[t]
\centering
\caption{Open-source response-envelope study. Entries give functionally successful data-envelope (D) implementations out of 15 trials. Columns specify local and sibling-resource patterns (F = flat, D = envelope); FN/DN indicate no sibling resource. $\Delta$ is the D-adoption difference between D-local and F-local assignments, with Holm-adjusted $p$-values across eight repository--agent strata. S = Sonnet-4-6; G6 = gpt-6-sol.}
\label{tab:osr-envelope}
\small
\setlength{\tabcolsep}{4pt}
\renewcommand{\arraystretch}{0.95}
\begin{tabular*}{\textwidth}{@{}l@{\extracolsep{\fill}}lrrrrrr@{}}
\toprule
 & & \multicolumn{2}{c}{Local flat} & \multicolumn{2}{c}{Local envelope} & & \\
\cmidrule(lr){3-4}\cmidrule(lr){5-6}
Repository & Agent & FF/FN & FD & DF/DN & DD & $\Delta$ & $p$ \\
\midrule
RealWorld & S  & 0/15 & 0/15 & 3/15 & 6/15 & 0.30 & 0.0078 \\
RealWorld & G6 & 0/15 & 0/15 & 12/15 & 11/15 & 0.77 & $1.2\times10^{-6}$ \\
Go REST & S  & 0/15 & 0/15 & 0/15 & 0/15 & 0.00 & 1 \\
Go REST & G6 & 0/15 & 0/15 & 15/15 & 15/15 & 1.00 & $1.5\times10^{-8}$ \\
ExpressJS & S  & 0/15 & 0/15 & 15/15 & 15/15 & 1.00 & $1.5\times10^{-8}$ \\
ExpressJS & G6 & 0/15 & 0/15 & 15/15 & 14/15 & 0.97 & $2.2\times10^{-8}$ \\
Clean & S  & 0/15 & --- & 15/15 & --- & 1.00 & $2.4\times10^{-4}$ \\
Clean & G6 & 0/15 & --- & 12/15 & --- & 0.80 & 0.0015 \\
\bottomrule
\end{tabular*}
\end{table}

\runhead{Error Handling.}
Agents follow the local pattern without exception. Of the 960 trials, 948 pass the hidden tests (98.8\%; Table~\ref{tab:osr-qualified}), and all 945 classifiable successes follow the local pattern (Wilson 95\% CI 99.6--100\%). In every repository--agent stratum, PP implementations propagate errors and DD implementations construct responses directly (16 Fisher tests, all $p<0.001$; Mantel--Haenszel $\chi^2\ge106.2$, $p<10^{-9}$ per agent), whereas changing only the sibling-resource pattern has no effect (all 16 Fisher tests, $p=1.0$).

\runhead{Response Envelope.}
Local precedent again dominates, though less uniformly (Table~\ref{tab:osr-envelope}). Of the 420 trials, 417 pass the hidden tests. With flat local handlers, no implementation adopts the envelope (0/207); with enveloped local handlers, 148 of 210 do (70\%). \texttt{gpt-6-sol} adopts the envelope in 11--15 of 15 runs in every repository, whereas Sonnet adopts it in 15/15 runs in \texttt{expressjs-api} and \texttt{fastapi-clean} but in none of 30 runs in \texttt{go-rest}. The local contrast is significant in 7 of 8 strata (differences 0.30--1.00), and none of the 12 sibling contrasts is significant. As in the controlled repository, response envelope is the choice that local precedent determines least completely.

\runhead{Explicit Guidance.}
A rule in the native instruction file redirects agents in real repositories. Without guidance, no run adopts the envelope (0/120), consistent with the 0/207 observed under flat local handlers in the response-envelope study above. With the rule, 117 of 120 runs both pass the hidden tests and adopt the envelope (97.5\%; Table~\ref{tab:osr-guidance}), an improvement of 80--100 percentage points in every repository--agent stratum (all eight Holm-adjusted $p=4.9\times10^{-4}$, the floor of the exact test with 15 blocks). \texttt{gpt-6-sol} follows the rule in 60 of 60 runs and Sonnet in 57 of 60. All three failures occur with Sonnet in \texttt{fastapi-realworld}: two runs keep a flat response and one adopts the envelope but fails a functional test, so raw target-shape adoption is 118/120 and functional success is 119/120. 

\rqanswer{Summary of the open-source study}{Across four open-source repositories spanning three languages, agents follow local precedent in all 945 functionally successful, classifiable error-handling implementations and 355/417 (85\%) functionally successful response-envelope implementations. Sibling-resource patterns have no detectable effect on either choice. An explicit rule in the harness's native instruction file redirects implementation choices despite conflicting local precedent, achieving 117/120 joint successes compared with 0/120 without guidance.
\implication The strong influence of local precedent and the effectiveness of explicit priority rules extend to the studied open-source repositories. However, adherence to local precedent varies across implementation choices, repositories, and agents.}

\section{Threats to Validity}\label{sec:threats}
\vspace{.3em}
\noindent{\bf Construct Validity.}
Our local--sibling comparison jointly varies task relatedness and repository location, so it does not distinguish their individual effects. The experiments control which precedents are available, not which ones agents attend to. The zero-precedent condition measures task-specific choices without relevant endpoint examples rather than context-free model preferences. Native instruction-file experiments evaluate rules loaded by the harness, not autonomous discovery. Our classifiers capture selected implementation patterns, and the outcomes concern implementation choice, not maintainability or overall code quality.

\vspace{.3em}
\noindent{\bf Internal Validity.}
In the support-level experiments, example count and file assignment vary together. The separate placement analysis reveals sensitivity to file identity but does not explain the influence of \texttt{tasks\_list}. The zero-precedent and precedent conditions were collected under different execution settings. Technical interruptions required retries and disrupted scheduling in RQ4. An initial misconfigured Opus file-placement run was excluded and rerun; open-source outcome audits also led to revised classifications and recapture of incomplete patches. These changes and the trial accounting are documented in the artifact. Finally, differences between agents cannot be attributed to models alone, as each agent combines a model with its harness.

\vspace{.3em}
\noindent{\bf External Validity.}
Fifteen repetitions per cell characterize variability under fixed conditions, not across tasks or repositories. The controlled experiments use one Flask API task family. The open-source study extends the local--sibling comparison to four repositories in three languages, but covers only two implementation choices, with explicit guidance evaluated for one. Support-level and file-placement effects remain untested beyond the controlled repository. The placement and guidance experiments cover selected settings, and the study evaluates agents from two providers.

\section{Related Work}

\vspace{.3em}
\noindent{\bf Coding Agents and Software Evolution.}
Code-generation benchmarks such as HumanEval and MBPP evaluate the functional correctness of standalone programs~\cite{chen2021evaluating,austin2021program}. SWE-bench moved evaluation to resolving real issues in existing repositories~\cite{jimenez2024swebench}, and agents such as SWE-agent~\cite{yang2024sweagent}, AutoCodeRover~\cite{zhang2024autocoderover}, OpenHands~\cite{wang2025openhands}, and Agentless~\cite{xia2025agentless} navigate, edit, and test codebases to resolve them. Recent work extends evaluation from isolated issues to the continuous evolution of a repository: SWE-Milestone~\cite{deng2026evoclaw} %
and SWE-EVO~\cite{le2026sweevobenchmarkingcodingagents} %
evaluate agents on multi-step, long-horizon evolution tasks, Needle in the Repo~\cite{zhu2026nitr} %
diagnoses maintainability failures in agent edits that pass functional tests, and Pinna et al.~\cite{pinna2026comparing} compare agents through pull-request acceptance. 

These studies ask whether agent changes succeed and remain maintainable. We ask which of several functionally acceptable implementations an agent selects, and what in the repository and task determines that choice.

\vspace{.3em}
\noindent{\bf Repository Context and Conventions.}
Software is repetitive and locally predictable~\cite{hindle2012naturalness,tu2014localness}, and project-specific conventions can be learned from existing code~\cite{allamanis2014natural}; LLM-generated code nonetheless deviates from human coding style~\cite{wang2025style}. Repository-level techniques retrieve or plan over cross-file context to improve generation and repository-wide edits~\cite{shrivastava2023repository,zhang2023repocoder,ding2023crosscodeeval,wu2024repoformer,li2025crossfile,bairi2024codeplan}, %
and repository instruction files such as \texttt{AGENTS.md} change agent behavior without generally improving task success~\cite{gloaguen2026evaluating}. 

These approaches treat context as a means to correctness. We hold correctness fixed and vary the available context, isolating how local precedent, sibling-resource precedent, and explicit instructions each govern the selected implementation.

\vspace{.3em}
\noindent{\bf Preferences, Conflicting Guidance, and Design Rationale.}
LLMs show systematic preferences in code generation, such as favoring Python and widely used libraries~\cite{twist2026preferences}, and their in-context behavior depends on the choice, order, and position of demonstrations~\cite{zhao2021calibrate,lu2022fantastically,fei2023mitigating,liu2024lost}. When context contradicts what a model has learned, LLMs may follow either source~\cite{xie2024adaptive,xu2024knowledge}, and instruction-hierarchy work trains models to rank instructions by their origin~\cite{wallace2024instruction}. We study the corresponding conflict in software evolution, where an agent's default, existing code, and new task intent compete, and measure which prevails. In software engineering, design decisions are first-class artifacts whose rationale is often lost~\cite{jansen2005software,liang2023implementation,zhao2024drminer,sharma2021rationale}. %
Our results show that agents act on recorded decisions without their rationale, and follow explicit priorities once these are stated.

\section{Conclusion}

As coding agents increasingly participate in software evolution, functional correctness alone cannot explain which implementation decisions they preserve or change. This study establishes implementation choice as a distinct dimension of agentic software engineering and yields three takeaways: (1) {repository precedent overrides agents' shared defaults}; (2) {where precedent appears can matter more than how often}, as local code dominates sibling resources and file placement alone can decide adoption; and (3) {new intent must be stated, not only demonstrated}, as written priority rules redirect agents more reliably than code examples.

Our findings call for evaluating not only whether agents produce correct changes, but how they reconcile past decisions with evolving intent. This requires design-decision traceability that preserves the rationale and priorities guiding future evolution.

\vspace{.3em}

\bibliographystyle{ACM-Reference-Format}
\bibliography{reference}

\end{document}